\documentclass[a4paper,fleqn]{cas-dc}
\usepackage[numbers]{natbib}
\usepackage{booktabs}
\usepackage{adjustbox}
\usepackage{longtable}
\usepackage{caption}
\usepackage{mathtools}
\usepackage{amssymb}
\usepackage{amsmath}
\usepackage{etoolbox}

\usepackage{array}   

\usepackage{doi}

\newlength{\myalignwidth}

\def\tsc#1{\csdef{#1}{\textsc{\lowercase{#1}}\xspace}}
\tsc{WGM}
\tsc{QE}
\tsc{EP}
\tsc{PMS}
\tsc{BEC}
\tsc{DE}

\begin{document}
\let\WriteBookmarks\relax
\def\floatpagepagefraction{1}
\def\textpagefraction{.001}
\let\printorcid\relax

\shorttitle{}

\shortauthors{Huang and Wang}

\title [mode = title]{Memory-based involution dilemma on square lattices}                      

\author[a]{Chaochao Huang}
\author[b]{Chaoqian Wang\corref{mycorrespondingauthor}}

\address[a]{School of Management, University of Science and Technology of China, Hefei 230026, China}
\address[b]{Department of Computational and Data Sciences, George Mason University, Fairfax, VA 22030, USA}

\cortext[mycorrespondingauthor]{Corresponding author\\
	Email address: \\
	chaochaohuang@mail.ustc.edu.cn~(C.~Huang);\\ CqWang814921147@outlook.com~(C.~Wang)}

\begin{abstract}
When involution affects individuals, their efforts do not augment resources but merely compete for limited resources. From entrance exams to company overtime, such efforts lead to unnecessary costs, undermining group welfare. Meanwhile, the individual advantages or collective disadvantages from this unnecessary effort may accumulate over time, such as the long-term validity of test scores. To identify the role of this memory factor, we propose a memory-based involution game model. In a more competitive environment, our findings suggest: (i) with scant social resources, increasing memory length curbs involution, (ii) with moderate resources, increasing memory length initially intensifies involution but later reduces it, and (iii) with abundant social resources, increasing memory length amplifies involution. Conversely, in a less competitive environment, involution consistently decreases with a larger memory length. Our research provides insights into mitigating involution by considering memory effects.
\end{abstract}




\maketitle

\section{Introduction}
The concept of the involution highlights a societal trend of futile competition~\cite{geertz1963agricultural}. It emphasizes that while the pool of resources remains fixed, players deploy their investments (efforts) serving merely as means to vie for these resources, without actually contributing to the augmentation of the resources~\cite{wang2021replicator,wang2022modeling}. Accordingly, the share each player receives is contingent upon their relative investments. 
In common examples of involution, including fixed-ratio admissions to universities and overtime in companies, the memory effect is a factor that cannot be ignored. The payoff due to an individual's efforts affect not only the current moment, but also the future. University admissions are still valid after a gap year. Some test scores for university admission, such as TOEFL and GRE, have long-term validity rather than instantly expendable. Working overtime at a company is more intuitive to interpret the memory effect---those employees who work overtime every day clearly accumulate better status. Having higher time-accumulation payoff as a result can lead other employees to imitate their overtime strategy.
The phenomenon of involution has been investigated in various domains, from psychology~\cite{kang2020review,liu2022have}, philosophy~\cite{huang2022analyzing}, to evolutionary game theory~\cite{wang2021replicator,wang2022modeling,wang2022involution,li2023involution,chaocheng2023involution}. In particular, evolutionary game theory provides a convenient research framework for involution, allowing quantitative factors such as memory to be introduced into the study.

Evolutionary game theory explains the evolution of cooperation by selfish human nature in social dilemmas~\cite{taylor1978evolutionary,nowak1998evolution,traulsen2006evolution,nowak2006five,perc2010coevolutionary}. Numerous seminal game models have been formulated to explore individual strategies, including the prisoner's dilemma~\cite{nowak1992evolutionary,szabo1998evolutionary,de2020interplay,bi2023based}, the snowdrift game~\cite{hauert2004spatial,wang2006memory,su2017spatial,shu2018memory,hu2021unfixed}, the stag hunt game~\cite{starnini2011coordination,dong2019memory,duong2020equilibrium} and the public goods game~\cite{szabo2002phase,liu2019effects,yang2019evolution,wang2021public,wang2022reversed,wang2023inertia}. Previous studies have been dedicated to introducing various mechanisms to promote and sustain cooperation. The classic mechanisms include reward~\cite{szolnoki2010reward,szolnoki2012evolutionary,shen2023evolutionary,sun2023combined}, punishment~\cite{szolnoki2011phase,szolnoki2011competition,wang2021tax,xiao2023evolution}, reputation~\cite{wang2017inferring,jian2021impact,shen2022high}, network topology~\cite{antonioni2017coevolution,civilini2023explosive,yang2023evaluation,wang2022decentralized}, exclusion~\cite{liu2019effects,quan2021comparison,quan2020social}, extortion~\cite{becks2019extortion,d2021human,chen2023outlearning}, and strategy updating rules~\cite{wang2022between,pi2022evolutionary,chen2023aspiration,wang2023zealous}. 

Among the aforementioned mechanisms, memory has been extensively applied to the research of classic game models~\cite{wang2006memory,shu2018memory,dong2019memory,liu2010memory,lu2018role,shu2019memory,li2023investigating}. Although in some cases memory may not necessarily promote cooperation~\cite{wang2014memory}, its realistic significance in the evolution of cooperation is undeniable. For instance, in human social interactions, reputation, grounded in memory, is pivotal in promoting cooperation and trust~\cite{xia2020effect}. In addition, memory-driven reciprocal behaviors serve as a fundamental reason for sustaining cooperation~\cite{shu2018memory,hilbe2017memory,li2022evolution,park2022cooperation,xia2023reputation}. In repeated interactions, individuals' past information (such as strategies and payoffs) can be observed, allowing for the establishment of reputation. Previous research incorporating memory have also explored the co-evolution~\cite{sun2020co} and robustness of cooperation~\cite{ren2014robustness}, while in currently research, memory continues to exhibit vitality~\cite{pi2022evolutionary,li2023investigating,duan2023evolutionary}.

Similar to many other social dilemma games, understanding and fostering the evolution of cooperation in the involution game is vital. Since the introduction of the involution dilemma into the field of evolutionary game theory by both replicator dynamics in well-mixed populations~\cite{wang2021replicator} and numerical simulations in structured populations~\cite{wang2022modeling}, some progress has been made. Ref.~\cite{wang2022involution} studied spatial and temporal heterogeneity in the social resources which players compete for. Ref.~\cite{li2023involution} explained the genesis of the social labor division from the perspective of involution games. Ref.~\cite{chaocheng2023involution} introduced the concept of ``lying flat'' and created a three-strategy system. Continuing from the previous studies, we seek to understand how the mechanism of memory affects the evolution of cooperation within the involution game. To specify, we propose a memory-based involution game model. In this model, the memory is represented by the average of the payoff from the previous $T$ steps. We also examine the interaction of several critical factors with memory length that affect the degree of involution, including the amounts of social resources, and the social temperature in resource distribution.

The organization of this paper is outlined as follows. In Section~\ref{sec_model}, we illustrate the memory-based involution game. In Section~\ref{sec_result}, we detail the primary outcomes derived from simulation experiments. Section~\ref{sec_concl} then offers a summation of the key discoveries.

\section{Model}\label{sec_model}
Consider involution games with agents located in the grids of a $G$-neighbor $L \times L$ square lattice with the periodic boundary condition. For a focal agent $i$, there is a group $\mathbb{N}(i)=\left\{j|\text{agent } j\text{ is a neighbor of agent }i \right \}$, and we assume $i \in \mathbb{N}(i)$. We suppose that at every time step $t$, each group is given fixed resources valued $M$, and each agent simultaneously chooses a strategy from cooperation (C) and defection (D): 
\begin{itemize}
	\item Cooperation (C): the agent incurs a standard cost $c$ and subsequently shares a certain proportion of resources.
	\item Defection (D): the agent incurs a competitive cost $d$ ($d > c$), providing an advantage in vying for a greater share of resources than a cooperator.
\end{itemize}
Based on their strategies, each agent secures a specific proportion of resources as described in Eq.~(\ref{eq1}),
\begin{equation}\label{eq1}
	 \pi_{t}(x)=\sum _{k \in \mathbb{N} (x)}\left (   \frac{\mathrm{e}^{\frac{o(x,t)}{\kappa _1} }}{\sum _{j \in \mathbb{N}(k) }\mathrm{e}^{\frac{o(j,t)}{\kappa _1} }}\cdot M-o(x,t) \right ) ,
\end{equation}
where $\pi_{t}(x)$ denotes the payoff of agent $x$ at time step $t$, and $o(x,t)$ signifies the associated cost, as illustrated in Eq.~(\ref{eq2}):
\begin{align} \label{eq2}
	o(x,t)=
	\begin{cases}
		c,   & \text{ the strategy of }x \text{ at time step }t \text{ is C}  , \\
		d,   & \text{ the strategy of }x \text{ at time step }t \text{ is D} . \\
	\end{cases}
\end{align}

When examining the coefficient preceding $M$ in Eq.~(\ref{eq1}), we recognize the application of the partition function. This aligns with the assumption that agents incurring higher costs have an advantage in resource allocation. Specifically, $\kappa_1$ ($\kappa_1>0$) represents the social temperature in resource distribution. Notably, a larger value of $\kappa_1$ indicates a more evenly distributed allocation in the system. It is also noteworthy that each agent is part of $G$ groups, receiving payoffs from $G$ games in one time step.

At the end of time step $t$ and before the beginning of the next time step $t+1$, agents evaluate their historical payoffs $P_t(x)$ within a finite memory length $T$, as shown in Eq.~(\ref{eq3}).
\begin{align} \label{eq3}
	P_t(x)=
	\begin{cases}
		\displaystyle{\frac{\sum _{\tau =1}^{t}\pi _{\tau}(x)}{t}},   & t \le T , \\
		\displaystyle{\frac{\sum _{\tau =t-T}^{t}\pi _{\tau}(x)}{T+1}},   & t>T . \\
	\end{cases}
\end{align}

Subsequently, agents adjust their strategies by imitating a randomly selected neighbor based on the difference in historical payoffs. We employ the Fermi function~\cite{szabo1998evolutionary}, as detailed in Eq.~(\ref{eq4}), to define the probability that agent $x$ will adopt the strategy of the selected agent $x_j$,
\begin{equation}\label{eq4}
	\text{Pro} \left\{ s_{x} \gets s_{x_{j}}\right\} = \left\{ 1+\exp \left ( \frac{P_{t}(x)-P_{t}(x_{j})}{\kappa _{2}} \right ) \right\} ^{-1},
\end{equation}
where the parameter $\kappa_{2}$ depicts the noise effect, representing how irrational factors influence agents' decision-making. As $\kappa_{2}$ approaches infinity, it results in neutral drift, while tending towards zero gives rise to deterministic pairwise comparison. Should agents be unsuccessful in strategy imitation, they retain their original strategies from time step $t$ for the ensuing game round. All agents synchronously update their strategies before commencing the subsequent game round. It is worth noting that when $T=0$ (indicating the memory-free scenario), the model degenerates to a foundational involution game model~\cite{wang2022modeling}.

Table~\ref{tb_value} summarizes the parameters (variables) and their interpretations.

\begin{table*}[width=\textwidth,pos=h]
\caption{List of the parameters (variables) and their interpretations.}\label{tb_value}
\begin{tabular*}{\tblwidth}{@{} LLL@{} }
\toprule
\multicolumn{1}{L}{Symbol} & 
Interpretation & Note
\\\midrule
$T$ &
The memory length. &
Input parameter; integer; $T\geq 0$; served for Eq.~(\ref{eq3}).
\\
$M$ &
The amount of social resources. &
Input parameter; $M>0$; served for Eq.~(\ref{eq1}).
\\
$c$ &
The cost of less effort. &
Input parameter; fix to $c=1$; served for Eq.~(\ref{eq1}).
\\
$d$ &
The cost of more effort. &
Input parameter; $d>c$; served for Eq.~(\ref{eq1}).
\\
$\kappa_{1}$ &
The social temperature of resource distribution. &
Input parameter; $\kappa_{1}>0$; served for Eq.~(\ref{eq1}).
\\
$\kappa_{2}$ &
The social temperature in updating strategies. &
Input parameter; fix to $\kappa_{2}=0.1$; served for Eq.~(\ref{eq4}).
\\
$o(x,t)$ &
The cost of agent $x$ at time step $t$. &
Auxiliary variable; obtained by Eq.~(\ref{eq2}); served for Eq.~(\ref{eq1}).
\\
$\pi_t(x)$ &
The payoff of agent $x$ at time step $t$. &
Auxiliary variable; obtained by Eq.~(\ref{eq1}); served for Eq.~(\ref{eq3}).
\\
$P_t(x)$ &
The averaged cumulative payoff of agent $x$ at time step $t$. &
Auxiliary variable; obtained by Eq.~(\ref{eq3}); served for Eq.~(\ref{eq4}).
\\
$f_D$ &
The fraction of defection, measuring involution level. &
Output variable; $0\leq f_D \leq 1$.
\\
\bottomrule
\end{tabular*}
\end{table*}

\section{Results and Discussion}\label{sec_result}
Define $f_D$ as the fraction of defectors among the $L \times L$ agents. Followed by Ref.~\cite{wang2022modeling}, we use $f_D$ as a proxy for the degree of involution. Because an increase in $f_D$ correlates with a rise in more efforts by agents and a corresponding decrease in the population's total payoff, thus such a pattern aligns with the core rationale of the concept of involution.
We conduct simulations centered on the research object $f_D$ with set parameters $L=100$, $G=5$, $c=1$ and $\kappa_{2}=0.1$. Upon setting the parameters, the game system evolves starting at time step $t=1$, where each agent has a 50\% probability of choosing either strategy C or D, until time step $t=10^3$. To mitigate data fluctuations in the experiment, we take the average of the last $300$ rounds of $f_D$ to represent a steady result for the given parameters.

\begin{figure*}
	\centering
	\makebox[\textwidth][c]{\includegraphics[width=\textwidth]{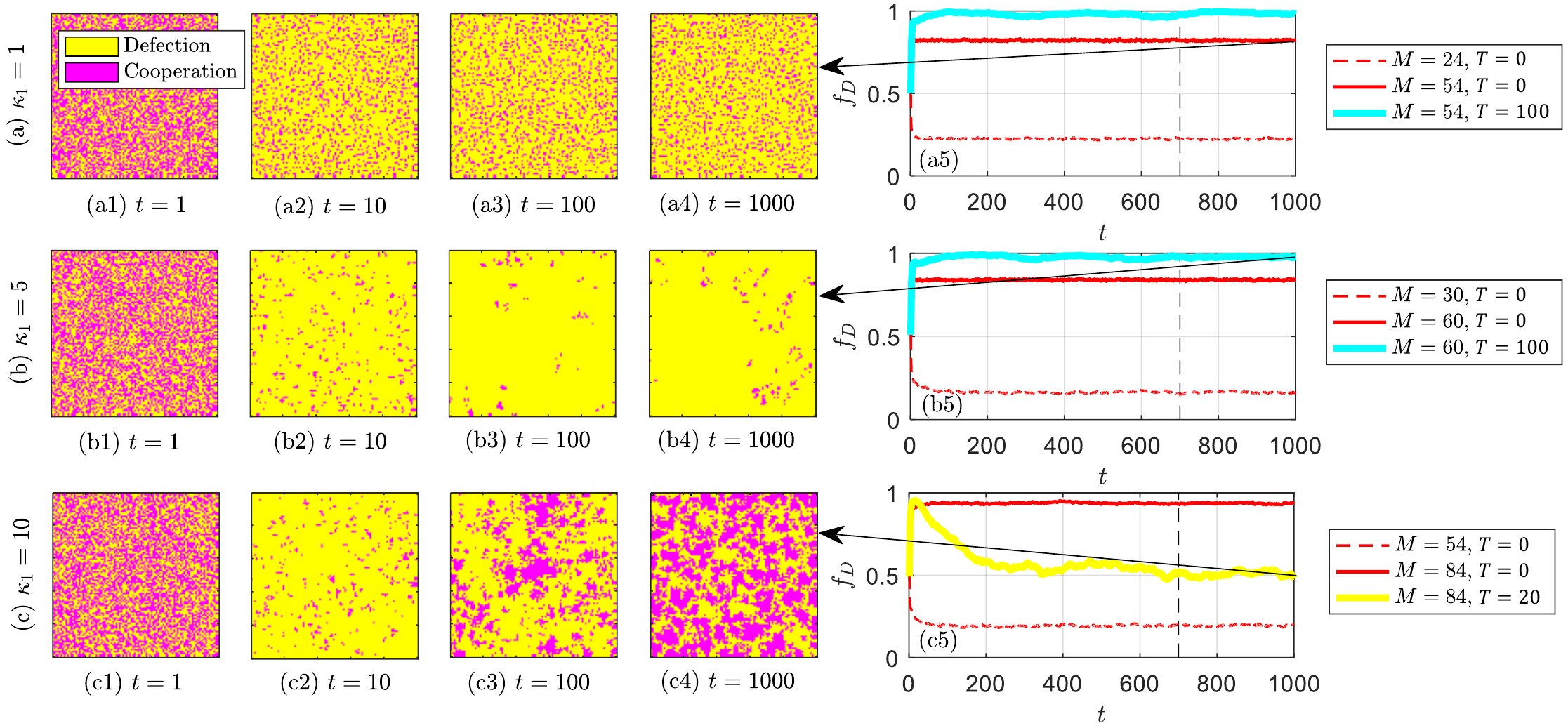}}
	\caption{The left $3 \times 4$ panels show the snapshots of the square lattice at $t=1$, $t=10$, $t=10^2$ and $t=10^3$. Yellow nodes indicate defective agents and purple nodes indicate cooperative agents. The values of ($M,T$) for panels (a1--a4), (b1--b4), and (c1--c4) are respectively ($54,0$), ($60,100$) and ($84,20$). See the subtitles for the value of $\kappa_1$. The right $3 \times 1$ panels (a5), (b5) and (c5) show the fraction of defectors $f_D$ as a function of time step $t$ at $\kappa_1=1,5,10$ correspondingly and different ($M,T$). Fixed parameter: $d=15$. }
	\label{Fig1}
\end{figure*}

Fig.~\ref{Fig1}(a5), (b5), and (c5) depict the fraction of defectors over time $t$ for varying values of $T$ and $M$. Comparing the curves in these three panels where $T=0$, it is evident that a higher value of $M$ results in a greater $f_D$ at the non-equilibrium statistical steady state. This observation aligns with findings from the previous work~\cite{wang2021replicator,wang2022modeling}. 
Next, we examine the influence of memory effects on the fraction of defectors. By observing the curves with $M=54$, $T=0$ and $M=54$, $T=100$ in Fig.~\ref{Fig1}(a5) and the curves with $M=60$, $T=0$ and $M=60$, $T=100$ in Fig.~\ref{Fig1}(b5), we notice that introducing memory can significantly increase $f_D$ when the system reaches stability. However, when comparing the curves with $M=84$, $T=0$ and $M=84$, $T=20$ in Fig.~\ref{Fig1}(c5) , we see that memory has the opposite effect, notably reducing $f_D$ in equilibrium. This indicates that the impact of memory on involution is not monotonic, but rather conditional, depending on the values of other parameters. 

We now turn to focus on the evolutionary details by the snapshots of the square lattice. The left $3 \times 4$ panels of Fig.~\ref{Fig1} show the snapshots of the square lattice at $t=1$, $t=10$, $t=10^2$ and $t=10^3$. Yellow nodes indicate defective agents and purple nodes indicate cooperative agents. In a memory-free context ($T=0$) with low resources and low social temperature in resource distribution ($M=54$, $\kappa_1=1$), the system rapidly stabilizes. In this steady state, cooperation and defection are evenly distributed (Fig.~\ref{Fig1}(a1--a4)). 
In a scenario with agents possessing large memory length ($T=100$) and moderate levels of both resources and social temperature in resource distribution ($M=60$, $\kappa_1=5$), the system predominantly disfavors cooperation (Fig.~\ref{Fig1}(b1--b4)).
In a context where agents have a moderate memory span ($T=20$) and both resources and social temperature are high ($M=84$, $\kappa_1=10$), the system takes longer to stabilize; however, cooperation is significantly favored. Cooperators amass in large numbers, resisting the invasion of defectors (Fig.~\ref{Fig1}(c1--c4)).

\begin{figure*}
	\centering
	\makebox[\textwidth][c]{\includegraphics[width=15cm]{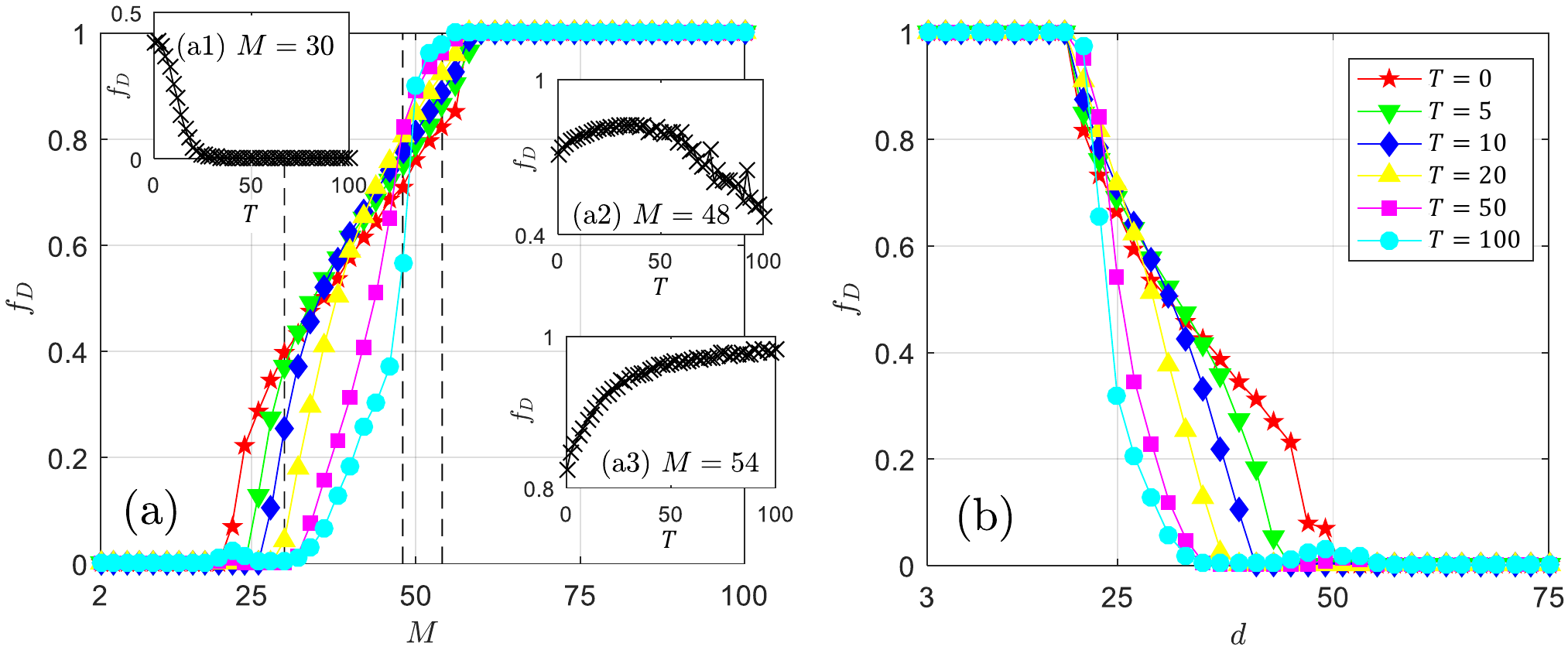}}
	\caption{(a) The fraction of defectors $f_D$ as a function of the amounts of resources $M$ under different memory length $T$ with fixed $\kappa_1=1$ and $d=15$. (a1--a3) The fraction of defectors $f_D$ as a function of memory length $T$ with $M=30$, $M=48$ and $M=54$, respectively. (b) The fraction of defectors $f_D$ as a function of competition cost $d$ under different memory length with fixed $M=76$, $\kappa_1=1$.}
	\label{Fig2}
\end{figure*}

Fig.~\ref{Fig2} presents the fraction of defectors $f_D$ as a function of resources $M$ and competition cost $d$ for various memory lengths $T$. We keep the social temperature constant at $\kappa_1=1$. In Fig.~\ref{Fig2}(a), the competition cost is set to $d=15$. We note that cooperators overwhelmingly dominate the network with small $M$ values. For varying $T$, a critical threshold $M^*(T,\kappa_1)|_{\kappa_1=1}$ emerges for the appearance of defectors, and this threshold rises as $T$ increases. Once $M$ surpasses the critical threshold $M^*(T,\kappa_1)|_{\kappa_1=1}$, defectors appear in the system, and $f_D$ continues to rise with increasing $M$, plateauing when $f_D=1$. 

Another intriguing observation from Fig.~\ref{Fig2}(a) is that for smaller values of $M$ (specifically, $M<30$), there is a decline in $f_D$ as memory length $T$ increases. Conversely, for larger $M$ values ($M>50$), $f_D$ rises with an increase in $T$. To explore a subtle relationship between the degree of involution $f_D$ and memory length $T$, we fix $M$ at $30,48$, and $50$ respectively in Fig.~\ref{Fig2}(a1--a3). In Fig.~\ref{Fig2}(a1), for a small $M$ value ($M=30$), we observe a consistent decline in $f_D$ with an increase in $T$, leveling off when $f_D=0$. Meanwhile, Fig.~\ref{Fig2}(a2) indicates that for a moderate $M$ value ($M=48$), the trend of $f_D$ is non-monotonic, initially rising and then falling with increased $T$. Lastly, Fig.~\ref{Fig2}(a3) depicts that for a larger $M$ value, $f_D$ consistently rises with $T$. 

In Fig.~\ref{Fig2}(b), where $M$ is set to 76, it is observed that with a low competition cost $d$, defectors primarily dominate the system. A critical threshold competition cost, $d^*(T,\kappa_1)|_{\kappa_1=1}$, for the emergence of cooperation is around the value of 20, without a notable correlation to the memory length $T$. In addition, $f_D$ shows a consistent decline with an increase in $d$, plateauing when $f_D=0$. For smaller $d$ values ($d<25$), the impact of memory on involution remains minimal. However, with larger $d$ values, an increase in $T$ leads to a significant decline in $f_D$.

\begin{figure*}
	\centering
	\makebox[\textwidth][c]{\includegraphics[width=16cm]{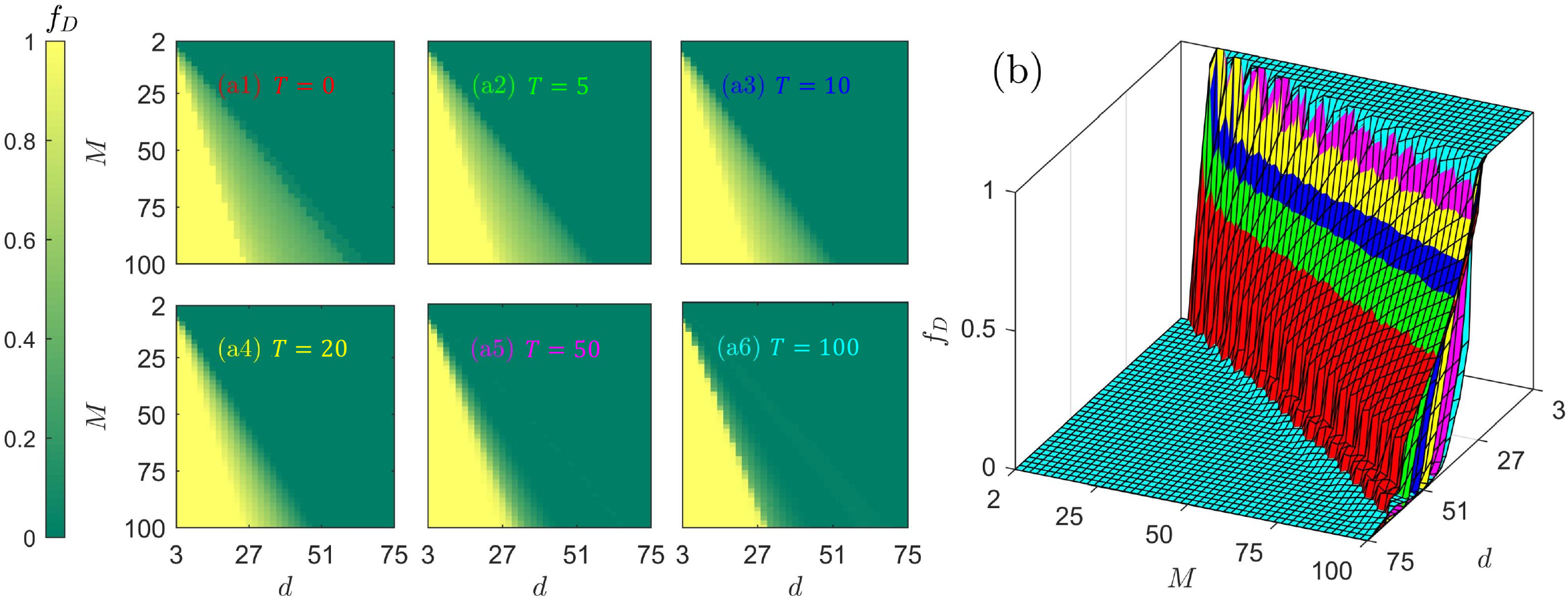}}
	\caption{(a1--a6) The fraction of defectors $f_D$ as a bivariate function of resources $M$ and competition cost $d$ under different memory length $T$ with fixed $\kappa_1=1$. (b) The three-dimensional version of panels (a1--a6).}
	\label{Fig3}
\end{figure*}

In Fig.~\ref{Fig3}, panels (a1--a6) present the fraction of defectors $f_D$ as a function of both resources $M$ and competition cost $d$, examined under various memory lengths $T$ ($T=0,5,10,20,50\text{ and } 100$, respectively), with $\kappa_1$ held constant at 1. Fig.~\ref{Fig3}(b) offers a three-dimensional representation of the data from panels (a1--a6). 
In Fig.~\ref{Fig3}, when $\kappa_1$ is set to a low value (i.e., $\kappa_1=1$), $f_D$ displays a monotonic variation with both $d$ and $M$: it reduces as $d$ increases and amplifies with $M$, regardless of the value of $T$, as highlighted in the yellow triangular region. The yellow triangular region slightly contracts as the memory length grows. This implies a mild suppression of involution by memory length when $\kappa_1$ is set to a low value.

\begin{figure*}
	\centering
	\makebox[\textwidth][c]{\includegraphics[width=15cm]{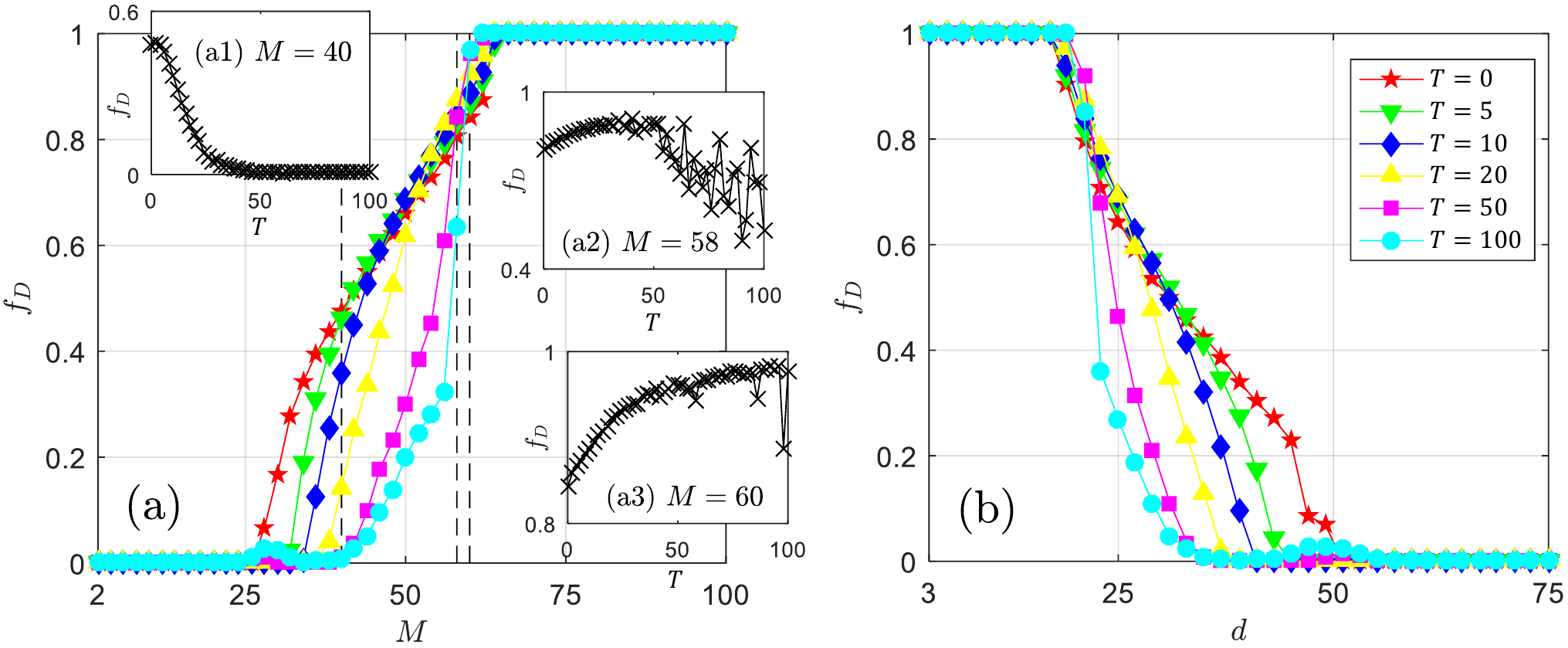}}
	\caption{(a) The fraction of defectors $f_D$ as a function of the amounts of resources $M$ under different memory length $T$ with fixed $\kappa_1=5$ and $d=15$. (a1--a3) The fraction of defectors $f_D$ as a function of memory length $T$ with $M=40,M=58$ and $M=60$ respectively. (b) The fraction of defectors $f_D$ as a function of competition cost $d$ under different memory length with fixed $M=76,\kappa_1=5$.}
	\label{Fig4}
\end{figure*}

Next, we aim to explore the impact of $\kappa_1$ on the involution degree $f_D$. We set $\kappa_1=5$ in the following simulation experiment. In Fig.~\ref{Fig4}, $f_D$ is presented as a function of resources $M$ and competition cost $d$ across different memory lengths $T$. The social temperature $\kappa_1$ remains constant at 5. As observed in Fig.~\ref{Fig4}(a) and akin to the simulation in Fig.~\ref{Fig2}(a), we fix the competition cost at $d=15$. First, compared to the results in Fig.~\ref{Fig2}, we find that when $\kappa_1$ increases from 1 to 5, there is no significant change in the critical resources required for the emergence of defectors, that is, $M^*(T,\kappa_1)|_{\kappa_1=1} \approx M^*(T,\kappa_1)|_{\kappa_1=5}$.
Second, we check the robustness of the prior conclusion. Specifically, for a smaller $M$ value ($M=40$), $f_D$ consistently decreases as $T$ increases. At a moderate $M$ ($M=58$), $f_D$ exhibits a non-monotonic behavior: it first rises and then falls with an increasing $T$. When $M$ is larger ($M=60$), $f_D$ consistently increases with $T$.

In Fig.~\ref{Fig4}(b), with fixed $\kappa_1=5$ and $M=76$, $f_D$ is presented as a function of competition cost $d$. The results align with those observed when $\kappa_1=1$ in Fig.~\ref{Fig2}(b), reaffirming our earlier conclusions. Specifically, at a lower competition cost $d$, the system is dominated by defectors. Cooperation emerges around a threshold competition cost of $d^*(T,\kappa_{1})|_{\kappa_{1}=5} \approx d^*(T,\kappa_{1})|_{\kappa_{1}=1}$, irrespective of the memory length $T$. Additionally, as $d$ increases, $f_D$ consistently decreases, stabilizing at $f_D=0$. For a smaller $d$, the impact of memory on involution is marginal. Yet, for a larger $d$, an uptick in $T$ results in a notable drop in $f_D$.

\begin{figure*}
	\centering
	\makebox[\textwidth][c]{\includegraphics[width=16cm]{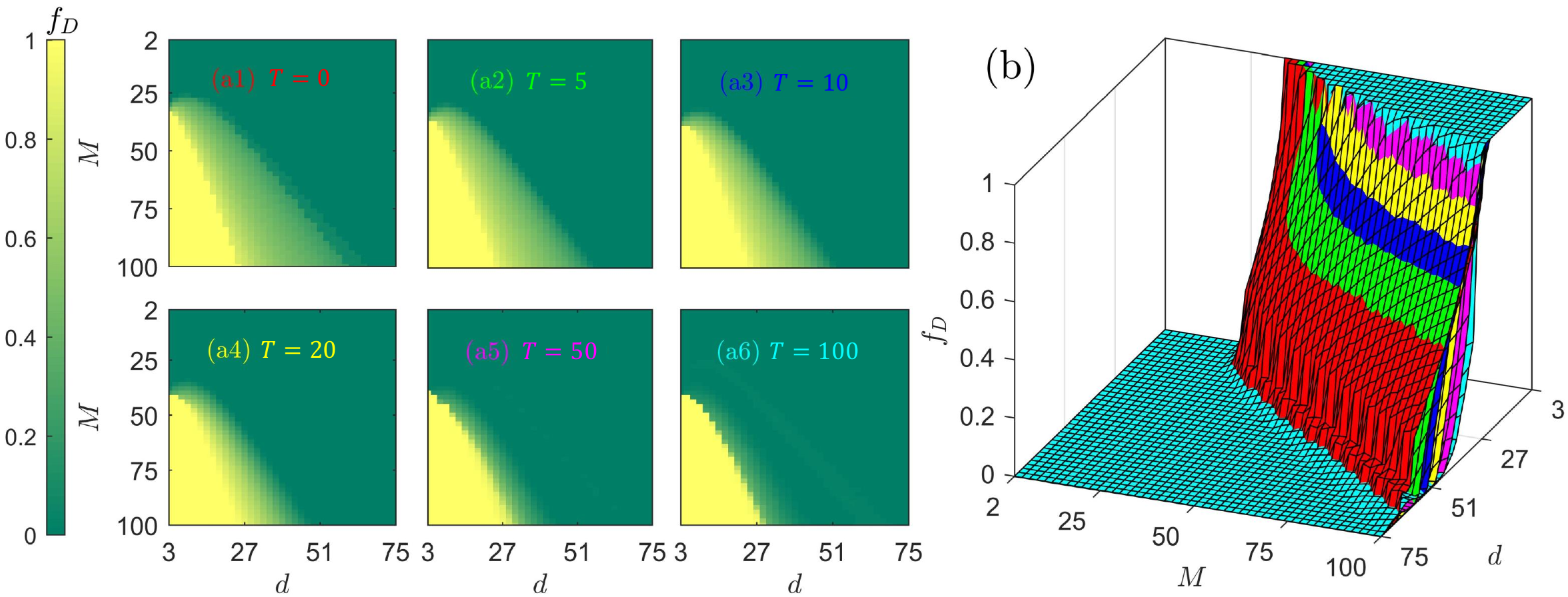}}
	\caption{(a1--a6) The fraction of defectors $f_D$ as a bivariate function of resources $M$ and competition cost $d$ under different memory length $T$ with fixed $\kappa_1=5$. (b) The three-dimensional version of panels (a1--a6).}
	\label{Fig5}
\end{figure*}

Fig.~\ref{Fig5}(a1--a6) illustrate $f_D$ as a function of $M$ and $d$ for varying memory length $T$ ($T=0,5,10,20,50$ and 100), keeping $\kappa_1$ constant at 5.
We observe that the yellow highlighted area in Fig.~\ref{Fig5}(a1--a6) is a subset of the yellow highlighted area in Fig.~\ref{Fig3}(a1--a6), implying that with an increase in $\kappa_1$, involution is generally suppressed. Another intriguing phenomenon is that, in contrast to the case when $\kappa_1=1$, for $\kappa_1=5$, when $M$ is relatively small ($M \approx 35$), $f_D$ no longer varies monotonically with $d$. Specifically, it increases initially with $d$ and then decreases. For a larger $M$, $f_D$ monotonically decreases as $d$ increases. Moreover, the monotonicity of $f_D$ with respect to $M$ is independent of the value of $d$, and it declines with an increase in $M$.

\begin{figure*}
	\centering
	\makebox[\textwidth][c]{\includegraphics[width=15cm]{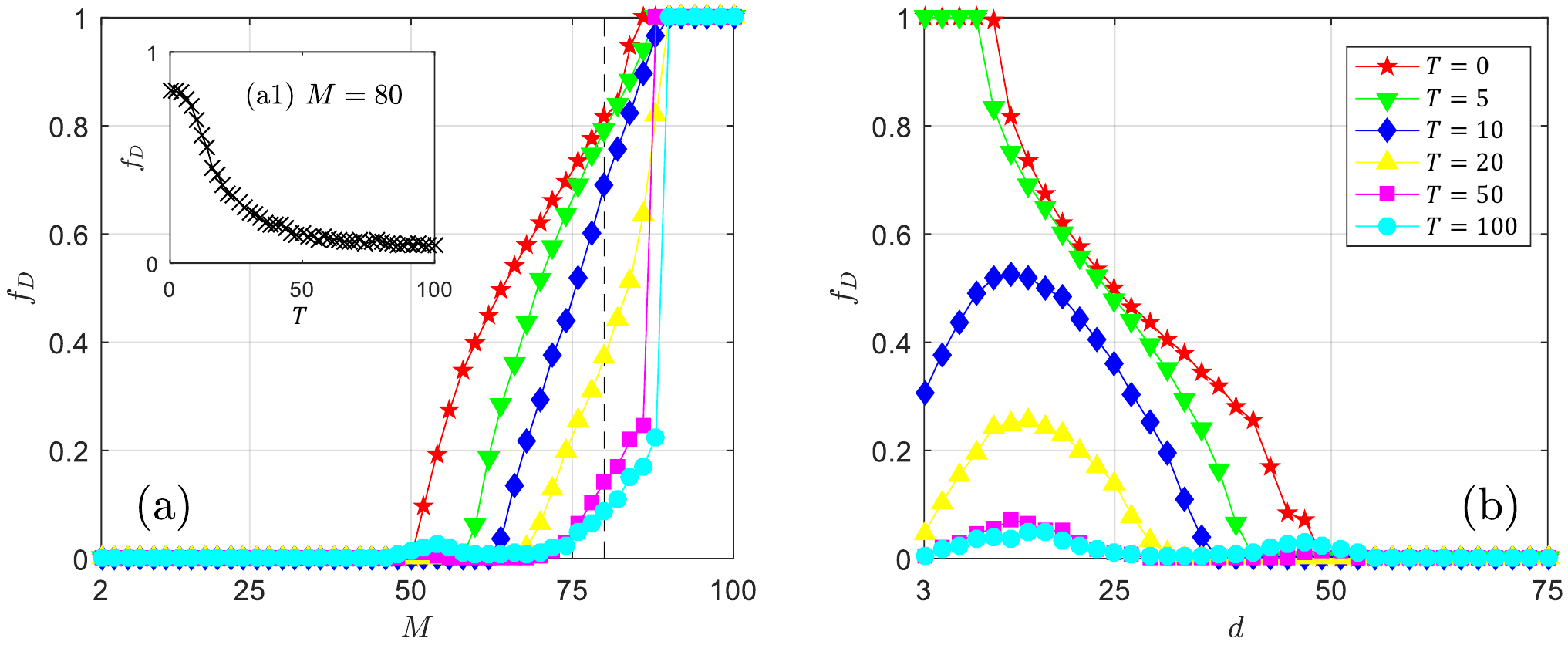}}
	\caption{(a) The fraction of defectors $f_D$ as a function of the amounts of resources $M$ under different memory length $T$ with fixed $\kappa_1=10$ and $d=15$. (a1) The fraction of defectors $f_D$ as a function of memory length $T$ with $M=80$. (b) The fraction of defectors $f_D$ as a function of competition cost $d$ under different memory length with fixed $M=76,\kappa_1=10$.}
	\label{Fig6}
\end{figure*}

Subsequently, we aim to check the robustness of the previous results when $\kappa_1$ is even larger ($\kappa_1=10$). Fig.~\ref{Fig6} illustrates $f_D$ as a function of $M$ and competition cost $d$ across different memory lengths $T$, maintaining $\kappa_1=10$ and $d=15$. We find that when $\kappa_1$ increases to 10, the qualitative result that $f_D$ increases with the rise of $M$ still holds. However, in contrast to Fig.~\ref{Fig2}(a) and Fig.~\ref{Fig4}(a), the monotonicity of $f_D$ with respect to $T$ is independent of $M$. Specifically, when $\kappa_1$ is sufficiently large, $f_D$ monotonically decreases with $T$ (See Fig.~\ref{Fig6}(a1) as an example).

Comparing Fig.~\ref{Fig6}(b) with Fig.~\ref{Fig2}(b) and~\ref{Fig4}(b), we observe that when $\kappa_1$ becomes large ($\kappa_1=10$), the qualitative relationship between $f_D$ and the competition cost $d$ changes (compared with $\kappa_1\in \{1,5\}$). Specifically, when $T$ is relatively small ($T \in \{0,5\}$), $f_D$ first remains constant at $f_D=1$ and then decreases to 0 as $d$ increases. When $T$ is larger ($T \in \{10,20,50,100 \}$), $f_D$ first rises and then falls with an increase in $d$, displaying a non-monotonic relationship. Moreover, unlike the scenarios with  $\kappa_1=1$ and $\kappa_1=5$, it is evident that even with a smaller $d$, an increase in $T$ significantly suppresses the degree of involution, rather than having no notable impact. Interestingly, for different values of $T$, $f_D$ peaks at nearly the same value of $d \approx 11$ in Fig.~\ref{Fig6}(b).

\begin{figure*}
	\centering
	\makebox[\textwidth][c]{\includegraphics[width=16cm]{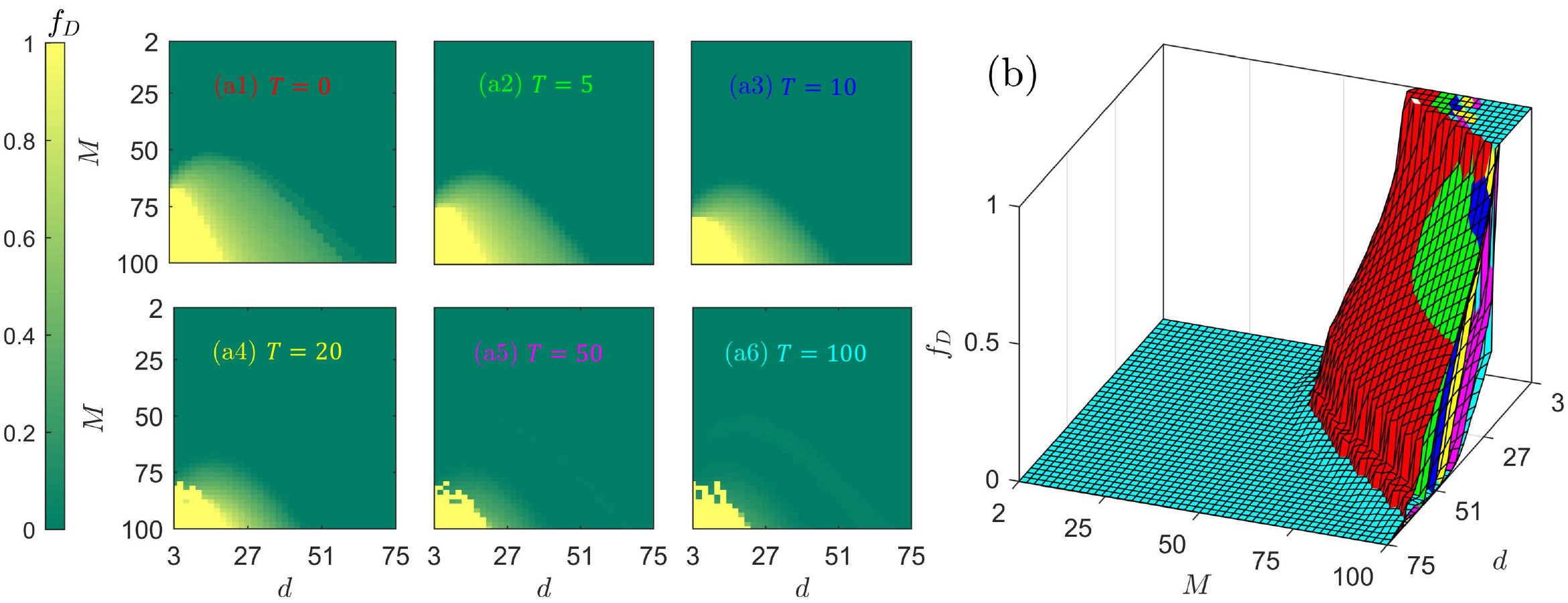}}
	\caption{(a1--a6) The fraction of defectors $f_D$ as a bivariate function of resources $M$ and competition cost $d$ under different memory length $T$ with fixed $\kappa_1=10$. (b) The three-dimensional version of panels (a1--a6).}
	\label{Fig7}
\end{figure*}

Panels (a1--a6) of Fig.~\ref{Fig7} depict $f_D$ as a function of $M$ and $d$ across diverse memory lengths $T$ ($T=0,5,10,20,50$ and $100$), with a consistent $\kappa_1$ value of 10.
Compared to the case with $\kappa_1=5$ (as shown in Fig.~\ref{Fig5}(a1--a6)), the previous results remain valid. Firstly, it demonstrates that an increase in $\kappa_1$ suppresses involution (with yellow highlighted area further contracting). Secondly, when $M$ is in a moderate range (e.g., $M\approx 70$), $f_D$ has a non-monotonic relationship with $d$, specifically, it first increases and then decreases with the rise of $d$. Lastly, an increase in $T$ suppresses involution. This effect is more pronounced when $T$ is smaller (for $T\in \{0,5,10,20\}$), and when $T$ is larger, the effect reaches saturation (e.g., for $T\in \{50,100\}$).

\section{Conclusion}\label{sec_concl}
When individuals suffer from involution, their efforts do not lead to an increase in resources, but merely serve as a means to compete each other for limited resources. Such meaningless effort brings redundant costs to the group and reduces social welfare. The objective of this research is to pinpoint the factors that influence the degree of involution, aiming to provide insights for its mitigation. Building on the spatial evolutionary game theory paradigm and drawing inspiration from prior research on the memory mechanism, we introduce a memory-based model to investigate the social dilemma of involution. We examine the model through four lenses: the amount of social resource, competition cost, social temperature in resource distribution, and memory length.

We assess the robustness of results from Ref.~\cite{wang2022modeling} across various memory lengths. Namely, increasing the number of social resources exacerbates involution. Higher social temperature tends to curb involution. Additionally, the degree of involution and competition cost exhibit a non-monotonic relationship. More specifically, as the amount of social resources falls within a moderate range, a rise in competition cost strengthens involution; in other cases, it hinders it. On this basis, we draw several novel results from the perspective of memory length.
\begin{enumerate}
	\item When the social temperature in resource distribution is low:
	\begin{itemize}
		\item For small amounts of social resources: Increased memory length curbs involution.
		\item For moderate amounts of social resources: A rise in memory length initially aggravates involution but later mitigates it.
		\item For larger amounts of social resources: Extended memory length amplifies involution.
	\end{itemize}
	\item When the social temperature in resource distribution is high: Involution consistently decreases with an expanding memory length.
	\item The suppressive impact of increasing memory length on involution is more pronounced at higher social temperatures in resource distribution.
\end{enumerate}

The memory length can be interpreted as the effective duration of an effort. For instance, as test scores for university admission have a validity period, policymakers might consider the longevity of various credentials. Re-evaluating the ``shelf-life'' of such resources can influence individuals' competition strategy and the overall societal stressors.


We consider future work from the following perspectives. First, one may incorporate moral preferences into the memory mechanism in the involution game. Recently, the moral preference hypothesis has emerged as a comprehensive model to elucidate various altruistic behaviors beyond mere cooperation. This hypothesis stresses individuals' inclination to adhere to personal norms---actions they consider morally right---beyond the monetary outcomes these actions yield. This framework is more effective in structuring cooperation across a broad spectrum of classical games and offers promising practical implications~\cite{capraro2021mathematical}. 
As a limitation of our current model, agents only consider their resource allocation and historical payoffs. However, in reality, agents might perceive eschewing involution as morally correct, since involution efforts do not enhance the amounts of social resources, whereas harm the collective interests of the population. Modeling the agents' payoff function when considering the trade-off between monetary gain (defection) and moral preference (cooperation) may mitigate involution.


Second, future work could integrate cutting-edge contexts of artificial intelligence (AI) into the framework of conditional strategies to investigate the memory-based involution game. Based on the memory mechanism, individuals can adopt conditional strategies in subsequent encounters, enabling either indirect or direct reciprocity~\cite{xia2023reputation}.
Similar to us, Refs.~\cite{han2021or,han2011intention} use the accumulated function to characterize memory effect. Contrastingly, Ref.~\cite{han2021or} considered a non-negligible cost of human-AI interaction. They introduced the ``trust-based'' strategy, which emphasizes randomly checking the co-player’s strategy once the number of cooperative interactions exceeds a certain threshold. Such strategies, unlike the traditional conditional strategies where individuals check their co-player's behavior in each round, help reduce the opportunity cost of verifying. On the other hand, in terms of information processing and integration, Ref.~\cite{han2011intention} proposed an ``intention recognition'' strategy based on a Bayesian network (BN), where agents can predict their opponent's strategy in the next round based on the historical information. In our model, agents learn opponents' strategies based on historical payoffs but lack the ability of prediction. We speculate that strategies akin to ``intention recognition'' could potentially alleviate involution. As revealed in Ref.~\cite{calvano2020artificial}, supracompetitive prices can sustain in oligopolistic price competitions (can be interpreted as an involution game) with AI-enabled algorithmic pricing.

\section*{CRediT authorship contribution statement}
Chaochao Huang: Writing--original draft, Result analysis. Chaoqian Wang: Model, Programming, Data filtering, Visualization, Writing--review \& editing.

\section*{Declaration of competing interest}
The authors declare that they have no known competing financial interests or personal relationships that could have appeared to influence the work reported in this paper.


\end{document}